\documentclass[epj]{svjour}
\usepackage{graphicx}
\usepackage{amsmath}
\usepackage{amssymb}
\usepackage[latin2]{inputenc}
\def\slaninafigdir{.}
\begin{document}
\title{%
Localization of eigenvectors in random graphs
}
\titlerunning{%
Localization  on random graphs
}%
\author{%
Franti\v{s}ek Slanina
\inst{1}%
\thanks{e-mail: {\tt slanina@fzu.cz
}
}}
\institute{
Institute of Physics,
 Academy of Sciences of the Czech Republic,
 Na~Slovance~2, CZ-18221~Praha,
Czech Republic
}
%
%
%
%
%
%
%
\abstract{
Using exact numerical diagonalization, we investigate localization in 
two classes of random matrices corresponding to random graphs. 
The first class comprises the adjacency matrices of Erd\H os-R\'enyi (ER) random graphs. 
The second one corresponds to random cubic graphs, with Gaussian random variables 
on the diagonal. We establish the position of the mobility edge, applying the 
finite-size analysis of the inverse participation ratio. The fraction of 
localized states is rather small on the ER graphs and decreases 
when the average degree increases. On the contrary, 
on cubic graphs the fraction of localized states is large and tends to 
$1$ when the strength of the disorder increases, implying that for sufficiently 
strong disorder all states are localized. The distribution of the
inverse participation ratio in localized phase has   finite width when
the system size tends to infinity and exhibits complicated multi-peak
structure.
We also confirm that the statistics of level spacings is Poissonian in
the localized regime, while for extended states it corresponds to
the Gaussian orthogonal ensemble.
}
\maketitle%
\section{Introduction}
After more than 50 years, Anderson localization \cite{anderson_58} remains
one of the most puzzling  problems of theoretical physics
\cite{abrahams_10}. Although many results have been accumulated
\cite{lee_ram_85,kra_mcki_93,eve_mir_08}, open 
questions remain even in the very basic issue of the definition of the
proper criterion of localization 
(as a single example, see e.g. \cite{rio_jit_las_sim_95}). From our viewpoint, however
subjective it might be, we can classify the approaches to the
phenomenon of localization into three big groups. In this introductory
sketch we shall emphasize the results concerning Bethe lattices, as
they are directly related to our work.

First, ``physical''
theories aim at grasping the essence without necessarily reaching the
mathematical rigor. A typical examples are the scaling theory \cite{abr_and_lic_ram_79},
 the self-consistent theory
\cite{vol_wol_80,vol_wol_92,suslov_95}, the approach based on parquet
diagrams \cite{jan_kol_05} and the approaches based on replica
\cite{wegner_79a} and supersymmetry \cite{efetov_83} methods. For our work, the
relevant sources are the results concerning localization on Bethe
lattice \cite{ab_an_tho_73,ab_tho_74,log_wol_85,ant_eco_77,gir_jon_80,efetov_90,mi_fyo_91}, 
where the exact self-consistent equation was formulated and
the localization threshold was computed. The phase diagram exhibits
extended states in the regime of weak 
disorder and energies sufficiently close to the band center. Otherwise
the states are localized. There is a well defined mobility edge,
separating extended states on one side from the localized states on
the other side. Although in principle we cannot exclude mixed regimes 
\cite{ku_sou_83}, 
in which localized and extended
states would coexist arbitrarily close to each other within a finite
interval, such a mixed regime was not yet observed.

Second, ``mathematical'' theories prove rigorously the localization
properties, but are limited to a few models where the known methods of
proof work. Still, there is a good deal of results available now, see
e. g. \cite{stollmann_01}. The result relevant for us is the proof of
localization in the Bethe lattice \cite{ku_sou_83,aiz_war_11}. 
However, the rigorous approaches work directly with infinite systems,
thus avoiding the difficulties in taking the thermodynamic limit. On
the other hand, it is the behavior of the system with increasing
size that is physically most interesting. Hence, the physical 
interpretation of the
rigorous results remains the matter of debate.

Third, one may resort to purely numerical computations, see e. g. 
\cite{markos_06} for electron localization or \cite{mon_gar_10} for
localization of acoustic waves. More sophisticated approaches rely on
the cavity approximation (which becomes exact on trees) and numerical
solution of thus obtained equation
\cite{ciz_bou_94,cav_gia_par_99,cil_gri_mar_par_ver_05,kuhn_08,met_ner_bol_10,bir_sem_tar_10,mon_gar_11}. 

The results on the localization in Bethe lattices bring the problem
close to the field of spectral theory \cite{chung_97} of random graphs
\cite{bollobas_85}, as many models
of random graphs are locally tree-like. Therefore, all local
properties of such random graphs should tend to Bethe lattice in
thermodynamic limit. Mathematically, spectra of random graphs are the
same thing as spectra of random sparse matrices. The latter were
studied in depth using various methods
\cite{rod_bra_88,rod_ded_90,fyo_mir_91,fyo_mir_91a,kho_rod_97,bi_mo_99,bau_gol_01,sem_cug_02,kuhn_08,rod_cas_kuh_tak_08}.  
Localization of eigenvectors was found both by exact numerical
diagonalization
\cite{evangelou_83,evangelou_92,eva_eco_92,ciz_bou_94,bi_mo_99,bir_sem_tar_10}  
and using the cavity method
\cite{cav_gia_par_99,cil_gri_mar_par_ver_05,kuhn_08,met_ner_bol_10}. Here
the study 
of random matrices touches again the problem of localization on a
Bethe lattice, as we mentioned above. 

Besides the academic interest in the localization phenomenon, numerous
examples of practical application of the ideas of localization can be
denonstrated, mainly in the area of complex networks
\cite{dor_gol_men_sam_03,sad_kal_hav_ber_05,car_and_sou_08,zhu_yan_yin_li_08} or in the 
field of the analysis of biological \cite{jal_sol_vat_li_10}
 and social networks \cite{gir_geo_she_09,sla_kon_10}.

We quoted several times the results showing the presence of
localization threshold for disordered Hamiltonians on Bethe
lattices. The fact is now confirmed by rigorous mathematic methods, as
well as physical arguments and numerical work on finite
samples. However, several problems remain. First, it is not quite
clear how the rigorous mathematic results should be translated to the
reality of physical experiments. Second, the Bethe lattice is pathological from
many points of view. Indeed, strictly speaking, in numerical studies
we work with a Cayley tree, rather than Bethe lattice. The difference
resides in the boundary conditions. In the  Cayley tree 
the
volume of ``surface'' sites is comparable to ``bulk'' sites, while the
negligibility of the former is the basis for the existence of basic
physical quantities, like the free-energy density. In the present work
we shall try to avoid the problem of surface by working with random
graphs. Our approach is based on the belief that in thermodynamic
limit the Bethe lattice and random graph results coincide. For a
mathematical justification, see \cite{bor_lel_08}.

In our previous work \cite{slanina_11} we showed that the cavity
approach, which may be 
considered as an approximation, coincides with the replica approach,
which is assumed to be exact, in thermodynamic limit.  The
variational formalism introduced in \cite{slanina_11} enables us to
consistently formulate approximations.

The present work is a continuation of that of
Ref. \cite{slanina_11}. First, we show how the formalism of
\cite{slanina_11} can be extended to study localization. The equations
found are in principle exact, but as soon as we resort to
approximations developed and used in Ref. \cite{slanina_11}, we find
that these approximations are insufficient to 
capture localization. Therefore, in the rest of the study
 we resort to exact numerical diagonalization
followed by finite-size scaling analysis.

\section{Cavity equations for localization}

Among the diverse criteria of localization, the most suitable for our
purposes is the behavior of the inverse participation ratio (IPR). 
Let $L$ be a $N\times N$ real symmetric matrix with eigenvalues
$\lambda_i$, $i=1,\ldots,N$ and corresponding normalized eigenvectors
$e_{j\lambda_i}$. We shall assume implicitly, that the matrix elements
of $L$ are random variables with properties described later.
The resolvent will be denoted $R(\zeta)=(\zeta-L)^{-1}$ and
its diagonal element $g_i(\zeta)=R_{ii}(\zeta)$ for
$\zeta\in\mathbb{C}\backslash\{\lambda_i;i=1,\ldots,N\}$.  
The IPR at $\lambda=\lambda_i$ for some $i$ is
\begin{equation}
\begin{split}
q^{-1}(\lambda)\,&=\sum_je_{j\lambda}^4=\\
&=\lim_{\varepsilon\to 0^+}
\frac{\varepsilon\,\sum_ig_i(\lambda+\mathrm{i}\varepsilon)\,g_i(\lambda-\mathrm{i}\varepsilon)
}{
\mathrm{Im}\sum_i  g_i(\lambda+\mathrm{i}\varepsilon)}\;.
\end{split}
\label{eq:defipr}
\end{equation}
For the proof of the latter equality, see \cite{eco_coh_72,met_ner_bol_10}.
The definition (\ref{eq:defipr}) applies for fixed system size
$N$. On the other hand, the question we ask in the analysis of
localization is, whether the states within a certain interval,
$\lambda\in I$, remain localized when $N\to\infty$ for all typical
realizations of the disorder. Therefore, we should define more
properly the average IPR in the interval $I$ as
\begin{equation}
\langle
  q^{-1}_I\rangle=\left\langle\frac{1}{N_I}\sum_{i:\lambda_i\in
    I}\sum_je_{j\lambda_i}^4
\right\rangle
\label{eq:defipraver}
\end{equation}
where 
 $\langle\ldots\rangle$ means averaging over the realizations of $L$ and
$N_I=\sum_{i:\lambda_i\in    I} 1$ 
is the number of eigenvalues within the interval $I$.
 Then,
if we find that $\langle
  q^{-1}_I\rangle \to 0$ as $N\to\infty$, the states in $I$ will be
considered extended, while non-zero limit would imply localization of
at least some of the states in the interval $I$.
We shall assume that the extended states, if they exist, are found
around the center of the spectrum, while localized states, if any,
should be expected at the upper and lower tails. More complicated
cases will not be treated here.
 The mobility edges
are then defined as numbers $z^-_\mathrm{mob}<z^+_\mathrm{mob}$ such
that 
\begin{equation}
\lim_{N\to\infty}\langle
  q^{-1}_I\rangle\;\;\left\{
\begin{array}{lll}
=0 &\text{  for any }& I\subset 
[z^-_\mathrm{mob},z^+_\mathrm{mob}]\\\\
>0 &\text{  for any }& I\subset 
(-\infty,z^-_\mathrm{mob})\\
&&\text{ or } I\subset (z^+_\mathrm{mob},\infty)\;.
\end{array}\right.
\label{eq:def-mobility-edge}
\end{equation}

Let us now sketch the formalism using the cavity method. It consists
in neglecting loops, so that it becomes exact on Bethe lattice, or on
any tree in general. We denote $g(\zeta)$
the diagonal element of the resolvent at the root of the tree. 
Following \cite{slanina_11} we introduce the generating functions
\begin{equation}
\begin{split}
\gamma(\omega)\,&=\left\langle e^{-\omega\,g(\zeta)} -1\right\rangle\\
\Gamma(\omega,\omega')\,&=\left\langle \left(e^{-\omega\,g(\zeta)} -1\right)\left(e^{-\omega'\,g(\zeta')} -1\right) \right\rangle\;.
\end{split}
\label{eq:generating-functions-twopar-def}
\end{equation}
The dependence on $\zeta$ and $\zeta'$ is assumed implicitly. We can
extract the linear and bilinear terms from the generating functions
 as
$\gamma(\omega)=\omega\,\big( s_1(\zeta)+O(\omega)\big)$ and
$\Gamma(\omega,\omega')=\omega\omega'\,\big(s_2(\zeta,\zeta')+O(\omega,\omega')\big)$. 
Hence we deduce the expression for the average IPR in the limit $N\to\infty$
\begin{equation}
q^{-1}(z)|_{N\to\infty}=\lim_{\varepsilon\to
  0^+}\frac{\varepsilon\,s_2(z+\mathrm{i}\varepsilon,z-\mathrm{i}\varepsilon)
}{
\mathrm{Im}\; s_1(z+\mathrm{i}\varepsilon)}
\;.
\label{eq:ipr-from-s1-s2}
\end{equation}
Strictly speaking, the expression (\ref{eq:ipr-from-s1-s2}) is
incorrect for two reasons. First, the order of the limits
$\varepsilon\to 0^+$ and $N\to\infty$ is reversed, because the cavity
approach works effectively with infinite $N$ from the very
beginning. Second, in (\ref{eq:ipr-from-s1-s2}) the average over
disorder  is performed separately in the numerator and in the
denominator, while if done properly, the averaging must involve the
fraction as a whole. Without entering into deep discussions, we assume
that neither of the two ``mistakes'' induce a fundamental fault into the
results. To support this assumption we can note, first, that also
Refs. \cite{met_ner_bol_10,kuh_mou_11} 
rely on the harmless exchange of the order of limits. Second, as for the
independent averaging of the numerator and denominator, it is
justified if we suppose
that $g_i(\lambda+\mathrm{i}\varepsilon)$ is a self-averaging
quantity, because in that case 
the disorder-average of the denominator is safely replaced by the sum $1/N\sum_i\bullet$.

If the degrees of the random graph are Poisson distributed, as is the
case for the Erd\H os-R\'enyi random graph, with average $\mu$, we
obtain, for the one-particle generating function $\gamma(\omega)$ a self-consistent
equation in the form
\begin{equation}
\begin{split}
&\gamma(\omega)=\sqrt{\omega}\int_0^\infty\, \frac{d\lambda}{\sqrt{\lambda}}\,
I_1(2\sqrt{\omega\lambda})\,\rho(\lambda)\\
&\rho(\omega)=e^{-\omega
  z+\mu\,\gamma(\omega)}\;.
\label{eq:pair-equations-gamma-rho}
\end{split}
\end{equation}
At this level, introduction of the auxiliary function $\rho(\omega)$
seems arbitrary, but it  acquires clear sense in the variational
approach developed in \cite{slanina_11}. 

For calculating IPR, the two-particle quantities are needed. Without
repeating the steps which lead to (\ref{eq:pair-equations-gamma-rho}),
we can write the equation for $\Gamma(\omega,\omega')$ as
\begin{equation}
\begin{split}
\Gamma(\omega,\omega')\,&=\sqrt{\omega\omega'}\int_0^\infty\,\frac{d\lambda}{\sqrt{\lambda}}\int_0^\infty\,\frac{d\lambda'}{\sqrt{\lambda'}}\,\times\\ 
&\times \,I_1(2\sqrt{\omega\lambda})\,I_1(2\sqrt{\omega'\lambda'})\,\rho(\lambda)\,\rho(\lambda')\times\\ 
&\times \,e^{\mu\,\Gamma(\lambda,\lambda')}\;.
\label{eq:twoparticle-gamma}
\end{split}
\end{equation}

Solving (\ref{eq:pair-equations-gamma-rho}) and 
(\ref{eq:twoparticle-gamma})
should in principle give full description of the localization
phenomenon. Note that the formalism used in \cite{ab_an_tho_73} and
\cite{cil_gri_mar_par_ver_05} should be a special case of
ours. Indeed, Refs. \cite{ab_an_tho_73,cil_gri_mar_par_ver_05} work with the
joint probability density for real and imaginary part of
$g(z+\mathrm{i}\varepsilon)$, which is equivalent to the joint
generating function for $g(z+\mathrm{i}\varepsilon)$ and
$g(z-\mathrm{i}\varepsilon)$. 

\begin{figure}[t]
\includegraphics[scale=0.85]{%
\slaninafigdir/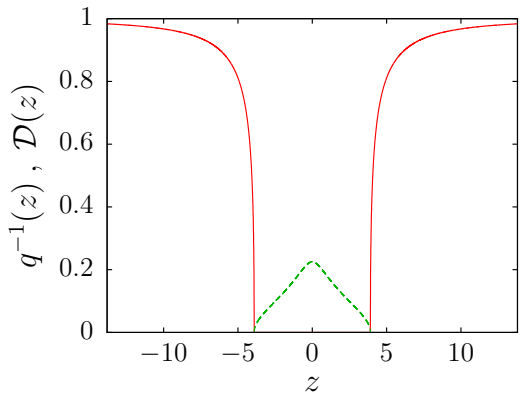}
\caption{Inverse participation ratio (solid line) and density of states
  (dashed line) calculated 
using the effective medium approximation, for Erd\H os-R\'enyi graph
with average degree $\mu$=3.}
\label{fig:randmat-ema-ipr-3}
\end{figure}

\begin{figure}[t]
\includegraphics[scale=0.85]{%
\slaninafigdir/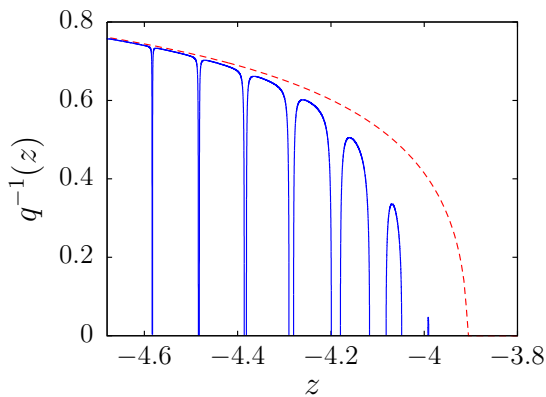}
\caption{Inverse participation ratio, for Erd\H os-R\'enyi graph
with average degree $\mu$=3 calculated 
using the effective medium approximation (dashed line) and
single-shell approximation (solid line).}
\label{fig:randmat-ema-ssa-ipr-3}
\end{figure}

Full solution of (\ref{eq:pair-equations-gamma-rho}) and 
(\ref{eq:twoparticle-gamma}) is not yet known. Approximative schemes
for solving (\ref{eq:pair-equations-gamma-rho}) were shown in
\cite{slanina_11}, partially repeating the older results of
\cite{bi_mo_99,sem_cug_02}. The simpler one of the approximations used
in \cite{slanina_11} is the effective-medium approximation (EMA),
which can be formulated as an ansatz
$\rho(\omega)=e^{\omega\,\sigma(\zeta)}$. For $\sigma(\zeta)$ we find the cubic equation
\begin{equation}
\sigma^3-\zeta\,\sigma^2+(\mu-1)\,\sigma+\zeta=0\;.
\label{eq:equation-for-sigma-ema}
\end{equation}
The density of states is non-zero only within the interval $ [ z_-,z_+
] $
where $\mathrm{Im}\,\sigma(z+\mathrm{i}\varepsilon)$ is non-zero in the limit
$\varepsilon\to 0^+$. Therefore, EMA exhibits sharp band edges, which
is wrong, because the true spectrum contains Lifschitz tails extending
arbitrarily far. Nevertheless, it is instructive to try to use EMA as a
starting point for approximative solution of the equation
(\ref{eq:twoparticle-gamma}). We insert in
(\ref{eq:twoparticle-gamma}) the functions $\rho(\lambda)$,
$\rho(\lambda')$, containing $\sigma(\zeta)$ obtained by solving
(\ref{eq:equation-for-sigma-ema}). Still, the resulting integral
equation for $\Gamma$ is not readily soluble, so we apply further
approximation, leaving only the lowest (bilinear) term in the
expansion of $\Gamma(\omega,\omega')$ and expanding $e^{\mu\Gamma}$ on
the right-hand side into series. This way we obtain an equation for
$s_2$ and the solution is then supplied into
(\ref{eq:ipr-from-s1-s2}). The IPR is then expressed through the
function $\sigma(z)$ for $z\in\mathbb{R}$. Finally we get
\begin{equation}
q^{-1}(z)=\frac{\big(3\sigma^2(z)-2z\sigma(z)+\mu-1\big)\sigma^4(z)
}{
\big(\sigma^2(z)-1\big)\big(\sigma^4(z)-\mu\big)}
\label{eq:ipr-through-sigma}
\end{equation}
for $z\in\mathbb{R}\backslash[z_-,z_+]$ and $q^{-1}(z)=0$ for
$z\in[z_-,z_+]$. The result is shown in Fig. \ref{fig:randmat-ema-ipr-3} for
$\mu=3$. We shall see later that this expression reflects qualitatively
well the behavior of IPR at the tails of the spectrum. However, the
result (\ref{eq:ipr-through-sigma}) is rather illusory, because
localization indicated by non-zero IPR occurs only in the areas where
density of states is strictly zero. The mobility edge coincides with
the band edge. Therefore, the fraction of
localized states is zero within such an approximation. 
We can try to improve the result applying the single-shell
approximation (SSA) introduced in  \cite{slanina_11}. Within this
approximation, we obtain for $\sigma$ the equation
\begin{equation}
z^2=
\mu+z\sigma+e^{-\mu}\sum_{l=1}^\infty
\frac{\mu^l}{(l-1)!}\,\frac{l}{z\sigma-l}\;.
\label{eq:equation-for-tau}
\end{equation}
As for the density
of states in the Lifschitz tail, SSA does give some improvement,
although severe artifacts of the approximation remain, namely the
spurious band gaps inside the Lifschitz tail. (See
Fig. \ref{fig:randmat-dos-3-tail} and 
\cite{slanina_11} for details.) In the same way as in EMA, we can take
the function $\sigma(\zeta)$ obtained in SSA, insert it into
(\ref{eq:twoparticle-gamma}) and expand $\Gamma(\omega,\omega')$ into
series. Thus, we obtain 
\begin{equation}
\begin{split}
q^{-1}(z)=&\,
\Bigg(
\frac{2}{1-e^{-\mu}\sum_{l=1}^\infty
\frac{\mu^l}{(l-1)!}\,\frac{l}{(z\sigma-l)^2}}
-\frac{\sigma}{z}
\Bigg)^{-1}\times\\
&\times\frac{\sigma^4(z)
}{
\sigma^4(z)-\mu}\;.
\end{split}
\label{eq:ipr-through-sigma-ssa}
\end{equation}
The result is shown in Fig.
\ref{fig:randmat-ema-ssa-ipr-3}. Contrary to EMA, the dependence of
the inverse participation ratio on eigenvalue is not monotonous, and
the ``interruptions'', where $q^{-1}(z)=0$ occur exactly at the
intervals where the density of states is non-zero. Therefore, SSA
suffers from the same flaw as EMA, that is the IPR is non-zero only if
density of states is zero. 
The conclusion
of this section is that analytical solution of
(\ref{eq:twoparticle-gamma}) would require more sophisticated methods
than those at our disposal.

In the rest of this work, we will rely on exact numerical
diagonalization results. However, the position of the band edge, as
found in EMA, will serve as a benchmark for the position of the
mobility edge and will be compared with numerical results.

\section{Localization in Erd\H os-R\'enyi graphs}

\begin{figure}[t]
\includegraphics[scale=0.85]{%
\slaninafigdir/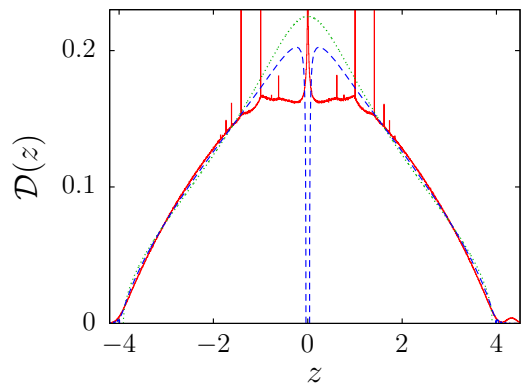}
\caption{The density of states for the adjacency matrix of the ER
  graph with $\mu=3$, $N=1000$, averaged over 115000 realizations (full
  line). For comparison, approximate results using EMA (dotted) and
  single-shell approximation of \cite{slanina_11} (dashed) are shown.}
\label{fig:randmat-dos-3}
\end{figure}

\begin{figure}[t]
\includegraphics[scale=0.85]{%
\slaninafigdir/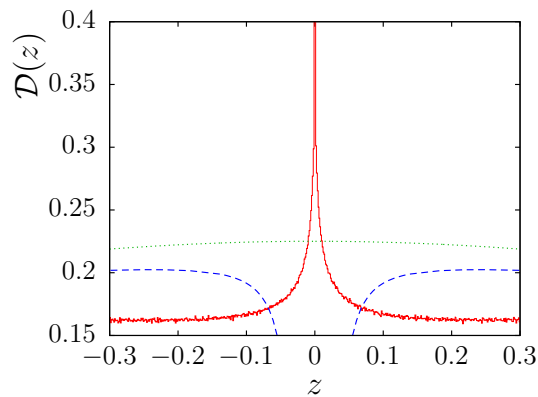}
\caption{Detail of the data of Fig. \ref{fig:randmat-dos-3}, showing
  the singularity at $z=0$.}
\label{fig:randmat-dos-3-detail-center}
\end{figure}

The first model we shall investigate is the adjacency matrix $L$ of
the Erd\H os-R\'enyi random graph. Apart from the fact that $L$ is
symmetric matrix with zero on the diagonal, the matrix elements are independent and equally
distributed. The  probability density for a single off-diagonal element is
\begin{equation}
\pi_1(x)=\Big(1-\frac{\mu}{N}\Big)\delta(x)+\frac{\mu}{N}\delta(x-1)\;.
\label{eq:distr-matrixelement}
\end{equation}
We investigated in depth the spectrum of $L$ in \cite{slanina_11}. In
Fig. \ref{fig:randmat-dos-3} we reproduce one of the results.
The density of states has a very complicated structure, with many
singularities and $\delta$-function components. For example, an 
acute, perhaps logarithmic, singularity resides at the center of the
spectrum, at $z=0$, as shown in
Fig. \ref{fig:randmat-dos-3-detail-center}. 
The theory exposed in Ref. \cite{bau_gol_01} could in principle bring an
explanation of that singularity, bud we did not perform the
calculations in this direction.

It is interesting to compare such suppression of localization in ER
graphs with the localization which occurs in weakly diluted systems,
where the localization is enhanced instead, by the mechanism of
maximum entropy walk \cite{bur_dud_luc_wac_09}. Indeed, on irregular graphs, for
example the common ER graph or a regular graph with a few edges
removed, the standard random walk does not posses maximum entropy. The
requirement of entropy maximization introduces a non-local constraint,
which, rather unexpectedly, favors localization.

\begin{figure}[t]
\includegraphics[scale=0.85]{%
\slaninafigdir/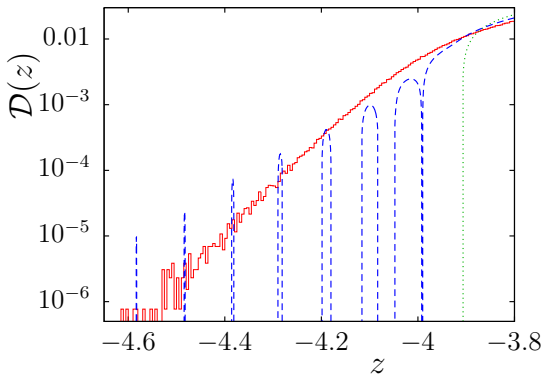}
\caption{Lower tail of the data of Fig. \ref{fig:randmat-dos-3}. Note
  that the single-shell approximation is superior to EMA in the
  Lifschitz-tail region, but still it is far from satisfactory.}
\label{fig:randmat-dos-3-tail}
\end{figure}

At the tails of the
spectrum, there is no sharp band edge, but a Lifschitz tail
develops. The asymptotic form of the Lifschitz tail is now well
established \cite{rod_bra_88,sem_cug_02,slanina_11} and our numerical
results can be seen in Fig. \ref{fig:randmat-dos-3-tail}.

\begin{figure}[t]
\includegraphics[scale=0.85]{%
\slaninafigdir/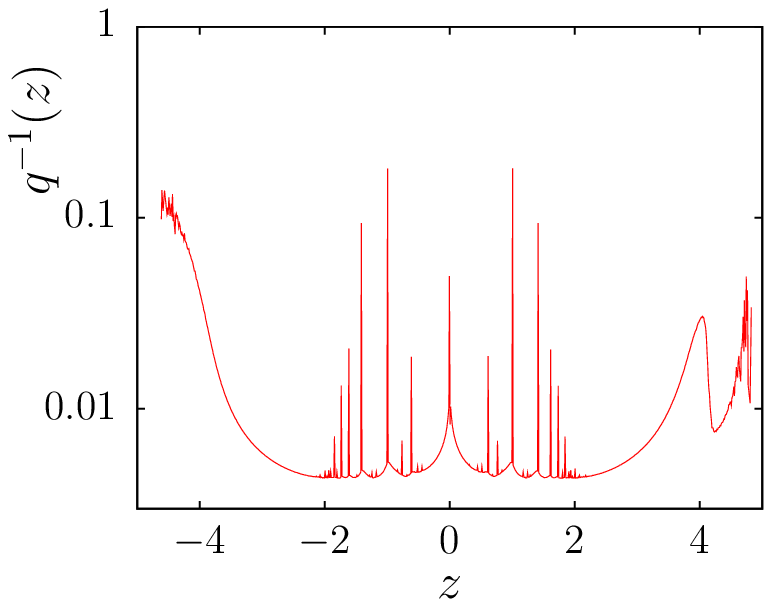}
\caption{Inverse participation ratio averaged over 115000 realizations,
  for ER graph with $\mu=3$ and $N=1000$.}
\label{fig:randmat-ipr-3}
\end{figure}

\begin{figure}[t]
\includegraphics[scale=0.85]{%
\slaninafigdir/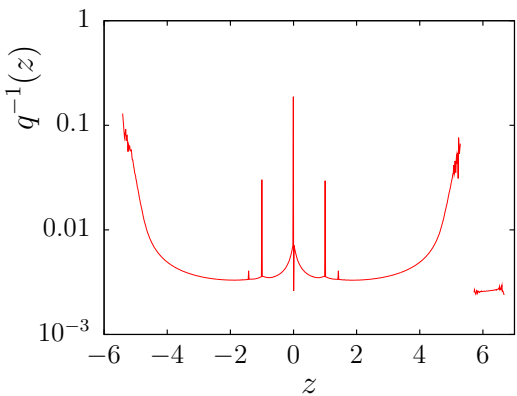}
\caption{Inverse participation ratio averaged over 65000 realizations,
  for ER graph with $\mu=5$ and $N=1000$.}
\label{fig:randmat-ipr-5}
\end{figure}

\begin{figure}[t]
\includegraphics[scale=0.85]{%
\slaninafigdir/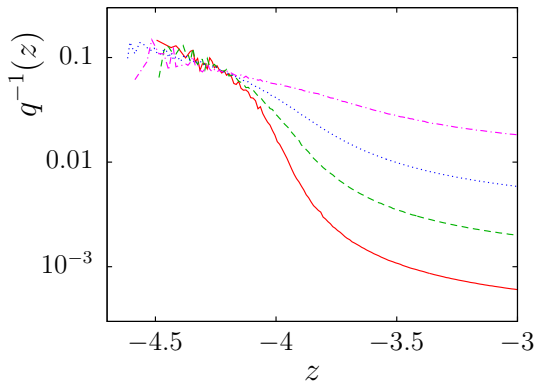}
\caption{Inverse participation ratio at the lower tail of the
  spectrum for ER graph with $\mu=3$. The system size is $N=10^4$
  (solid line), $3000$ (dashed line), $1000$ (dotted line), and $300$
  (dash-dotted line). The data are
 averaged over $900$, $10000$, $65000$ and $130000$ realizations,
 respectively.}
\label{fig:randmat-ipr-3-tail}
\end{figure}

\begin{figure}[t]
\includegraphics[scale=0.85]{%
\slaninafigdir/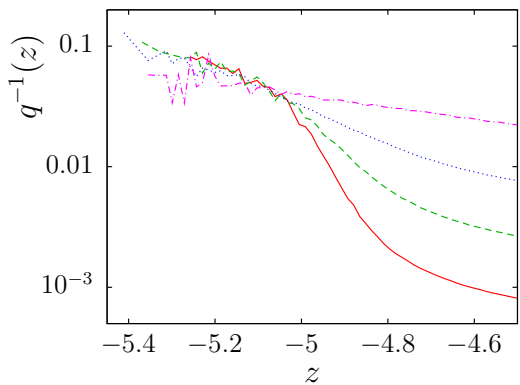}
\caption{Inverse participation ratio at the lower tail of the
  spectrum for ER graph with $\mu=5$. The system size is $N=10^4$
  (solid line), $3000$ (dashed line), $1000$ (dotted line), and $300$
  (dash-dotted line). The data are
 averaged over $800$, $5000$, $65000$ and $50000$ realizations,
 respectively.}
\label{fig:randmat-ipr-5-tail}
\end{figure}

It is just the Lifschitz tail where the localization is expected. To
have a first glance on that, we plot the IPR averaged over several
tens of thousand realizations. In Figs \ref{fig:randmat-ipr-3} and
\ref{fig:randmat-ipr-5} we show the results for  $N=1000$ and for
$\mu=3$ and $\mu=5$, respectively. Comparing the behavior of IPR with
the density of states, as shown in Fig. \ref{fig:randmat-dos-3}, we
observe the same complicated structure of singularities. Generally,
IPR is large at the tails, as well as close to the singularities in
the density of states. One would naively expect that localization
would occur in all regions where IPR is large, but it is true only in
the tails. As we stressed earlier, we must check the behavior of IPR
when $N$ grows. Close to the singularities, we found IPR large, but
consistently decreasing with increasing system size. On the contrary,
localization in Lifschitz tails is clearly visible, as indicated in
Figs \ref{fig:randmat-ipr-3-tail} and \ref{fig:randmat-ipr-5-tail}. In
the following, we
decided to work with the lower tail, because the upper tail is
somewhat obscured by the single maximum eigenvalue which behaves
differently than all the rest of the spectrum. We can see that below
certain value of $z$, the IPR is independent of $N$, within the range
of statistical errors, while above this value, IPR decreases with
$N$. 
We identify this value with the mobility edge. We shall describe the
method of extracting the mobility edge from the data in the next
section. Here we make only a few observations. 

First, this definition
of mobility edge is practical but it is not the only possible. Moreover,
there might be even some doubts of it. Indeed, above the mobility
edge the IPR should not only decrease with $N$, but decrease in a
specific manner, namely as $1/N$, otherwise the states cannot be
considered properly extended. Therefore, the alternative definition of
the mobility edge would be as follows. We declare the states
in the interval
$I$  extended, if $\langle q^{-1}_I\rangle\sim N^{-1}$ for
$N\to\infty$, otherwise the states are considered localized.
 The data from Figs
\ref{fig:randmat-ipr-3-tail} and \ref{fig:randmat-ipr-5-tail} indicate
that the mobility edge defined in the latter way would lie somewhat
higher than in the former. 
The difference may well be just a finite-size effect, but we
cannot exclude also the possibility that it reflects a real
phenomenon, namely presence of states which are neither properly
extended, nor exponentially localized. For example, the eigenvectors
can be characterized by tails decreasing slower than any exponential,
but on the level of knowledge provided by our numerical data 
this is a mere speculation. However, note that eigenvectors with
power-law tails do occur in certain models
\cite{del_sim_sou_84,del_sim_sou_85} and 
an interval of 
coexistence of extended and localized states was also hypothesized in
\cite{ku_sou_83}. 
In all the rest we shall stick to the former definition of the
mobility edge for purely practical
reasons. 

Second, comparing the IPR calculated using EMA and SSA, shown in
 Fig. \ref{fig:randmat-ema-ssa-ipr-3}
 with
numerical findings in Figs \ref{fig:randmat-ipr-3-tail} and
 \ref{fig:randmat-ipr-5-tail} , we observe a qualitative 
agreement. On the other hand, quantitatively, EMA and SSA give much
too high values of IPR. So, however defective EMA and SSA are 
with respect to localization,
they do provide a hint of how IPR should behave.

Third, the data suggest that IPR for infinite system 
approaches a non-zero limit when we
approach the mobility edge from the localized side. Because in extended
regime IPR is strictly zero for infinite system, IPR should exhibit a
discontinuity at the mobility edge. This confirms results obtained
earlier in \cite{fyo_mir_som_92} using supersymmetric method.

\begin{figure}[t]
\includegraphics[scale=0.85]{%
\slaninafigdir/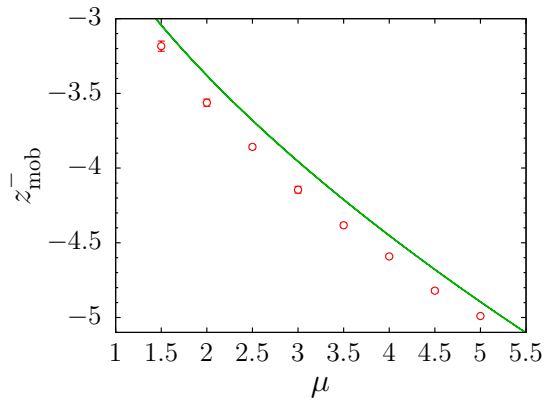}
\caption{Position of the mobility edge at the lower tail of the
  spectrum, for ER graph. Where not shown, error bars are smaller than
  the symbol size. The solid line is the band edge calculated in EMA.}
\label{fig:randmat-mobilityedge-lower}
\end{figure}

Let us continue with the analysis of
 our results. Having established the mobility edge,
we want to know how it depends on the average degree of the ER
graph. This dependence is shown in
Fig. \ref{fig:randmat-mobilityedge-lower}. For comparison, we show
also the position of the band edge, as found in EMA. We can see that
the mobility edge is slightly below the EMA band edge, but the two
quantities share a common trend. Therefore, the EMA band edge can
serve as a useful zeroth approximation for the line of separation
between localized and extended states. This criterion was used,
without further justification, in the context of diffusion models for
biological evolution \cite{kol_sla_03}. 

\begin{figure}[t]
\includegraphics[scale=0.85]{%
\slaninafigdir/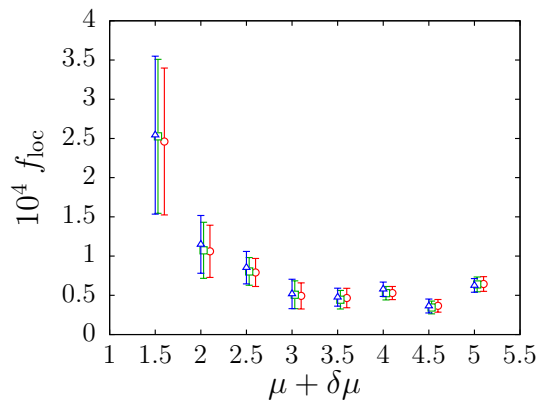}
\caption{Fraction of states below the lower mobility edge, for ER graph. We
  compare the results for $N=10000$ (triangles), $3000$ (squares), and
  $1000$ (circles). For better visibility, the points are slightly shifted
  rightwards by $\delta\mu=0$, $0.03$, and $0.1$ for $N=10000$,
  $3000$, and $1000$, respectively.}
\label{fig:randmat-fraction-localised}
\end{figure}

In order to see quantitatively, how much globally relevant is the
localization phenomenon, we measure the fraction of eigenvalues below
the mobility edge 
\begin{equation}
f_\mathrm{loc}=\left\langle\frac{1}{N}
\sum_{i:\lambda_i < z^-_\mathrm{mob}}1\right\rangle\;.
\label{eq:frac-loc-definition}
\end{equation}
Supposing that the spectrum is mirror-symmetric,
as it should be in the limit $N\to\infty$, the total fraction of
localized states is $2f_\mathrm{loc}$. We can see the results in
Fig.~\ref{fig:randmat-fraction-localised}. The first thing to note is
that the results are practically independent of system size, so we can
safely claim that they represent the fraction of localized eigenvalues
for infinite system. The fraction decays with average degree $\mu$,
until it saturates around $\mu\simeq 3$ at a value close to
$f_\mathrm{loc}\simeq 0.5\cdot 10^{-4}$. It is suposed that this
fraction should drop to zero in the limit $N\to\infty$, because it is
known that  all states are extended in an ER graph, on condition that
$\mu\to\infty$ simultaneously with $N\to\infty$
\cite{erd_kno_yau_yin_11}. The
numerical procedure does not enable us to work with large enough $N$
to see that explicitly. Therefore, we consider the saturation a
finite-size effect.

\section{Localization in random cubic graphs}

\subsection{Diagonal disorder}

The second family of graphs investigated here are the random cubic graphs,
i. e. random graphs satisfying the only constraint that the degree of
all vertices is equal to $3$. 
We decided to study this family as a kind of direct opposite of the ER
graph. In ER graph, the properties are mostly due to inhomogeneity in
the degrees. In cubic graph all degrees are equal. In ER graph, there
is no diagonal disorder. In cubic graph, the relevant disorder is
only on the diagonal. Of course, one can study also models which
interpolate the two extremes, but we shall not do that in the present
work. We shall rather compare the differences between the extremes.

The off-diagonal elements of the 
matrix $L$ to study are identical to the adjacency matrix of the
graph, while diagonal elements of $L$ are independent Gaussian random
variables, with probability density 
\begin{equation}
\pi_\mathrm{diag}(L_{ii})=\frac{1}{\sqrt{2\pi}\,\eta}\exp\left(-\frac{L_{ii}^2}{2\eta^2}\right)
\label{eq:distr-diagonal}
\end{equation}

In thermodynamic limit the
local topology of the graph is identical to the Bethe lattice with
coordination number $3$ and the randomness of the structure, i. e. the
off-diagonal disorder, must be
irrelevant, as long as we investigate local properties of the graph
and its size goes to infinity. For example, the density of states for
the random graph with $\eta=0$ must
approach a non-random function identical to the well-known density of states 
of the Bethe lattice
\begin{equation}
\mathcal{D}_\mathrm{Bethe}(z)=\frac{3}{2\pi}\frac{\sqrt{8-z^2}}{9-z^2}\;.
\label{eq:distr-diagonal}
\end{equation}
 The non-trivial ingredient is the randomness in
diagonal elements of the matrix $L$ and this is the feature which
leads to localization here. The situation is somewhat complementary to
the ER case investigated in the last section. In ER graphs,
localization is due to off-diagonal disorder, while here the diagonal
disorder is responsible.

\begin{figure}[t]
\includegraphics[scale=0.85]{%
\slaninafigdir/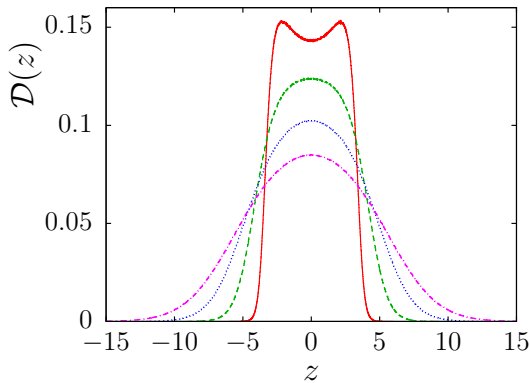}
\caption{Density of states for random cubic graph with diagonal
  disorder, for $N=1000$. The disorder strength is $\eta=1$ (solid
  line), $2$ (dashed line), $3$ (dotted line), and $4$ (dash-dotted
  line). The data are averaged over $40000$ realizations.}
\label{fig:randcubic-dos-1-2-3-4}
\end{figure}

\begin{figure}[t]
\includegraphics[scale=0.85]{%
\slaninafigdir/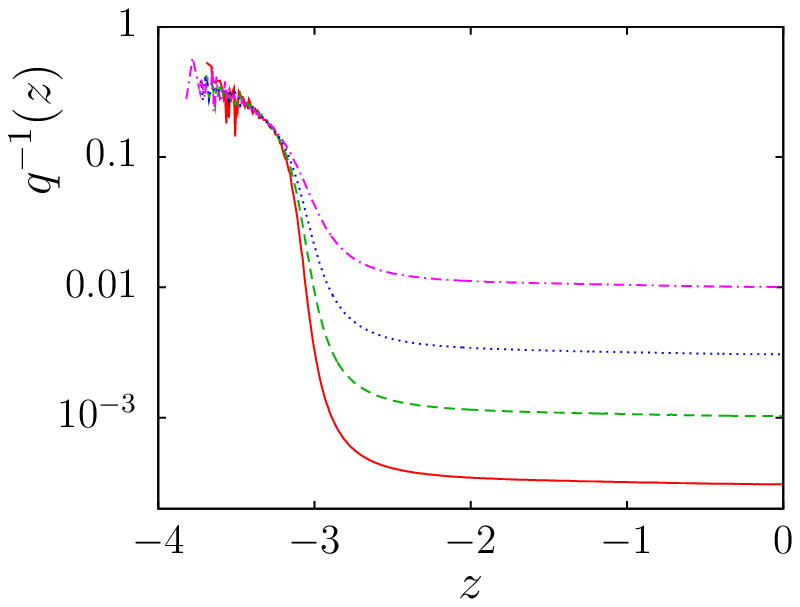}
\caption{Inverse participation ratio at the lower tail of the spectrum
for random cubic graph with disorder strength $\eta=0.5$. The system size is $N=10^4$
  (solid line), $3000$ (dashed line), $1000$ (dotted line), and $300$
  (dash-dotted line). The data are averaged over $550$, $11000$,
$65000$, and $160000$ realizations, respectively.}
\label{fig:randcubic-ipr-0.5-tail}
\end{figure}

\begin{figure}[t]
\includegraphics[scale=0.85]{%
\slaninafigdir/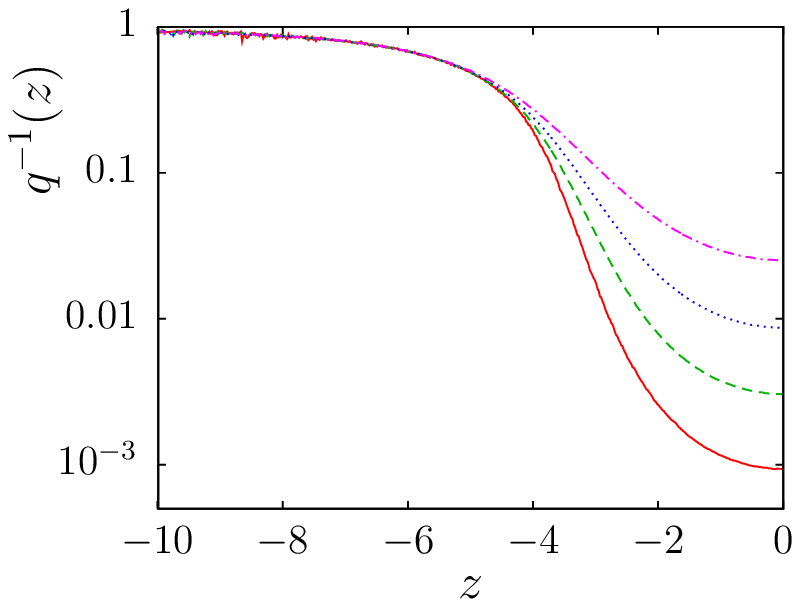}
\caption{Inverse participation ratio at the lower tail of the spectrum
for random cubic graph with disorder strength $\eta=2$. The system size is $N=10^4$
  (solid line), $3000$ (dashed line), $1000$ (dotted line), and $300$
  (dash-dotted line). The data are averaged over $610$, $10000$,
$65000$, and $160000$ realizations, respectively.}
\label{fig:randcubic-ipr-2-tail}
\end{figure}

\begin{figure}[t]
\includegraphics[scale=0.85]{%
\slaninafigdir/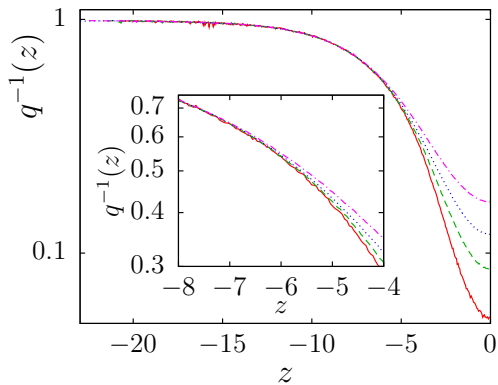}
\caption{Inverse participation ratio at the lower tail of the spectrum
for random cubic graph with disorder strength $\eta=4$. The system size is $N=10^4$
  (solid line), $3000$ (dashed line), $1000$ (dotted line), and $300$
  (dash-dotted line). The data are averaged over $550$, $10000$,
$65000$, and $160000$ realizations, respectively. In the inset, detail
of the data illustrating the difficulty to establish the mobility edge precisely.}
\label{fig:randcubic-ipr-4-tail}
\end{figure}

\subsection{Mobility edge}

\begin{figure}[t]
\includegraphics[scale=0.85]{%
\slaninafigdir/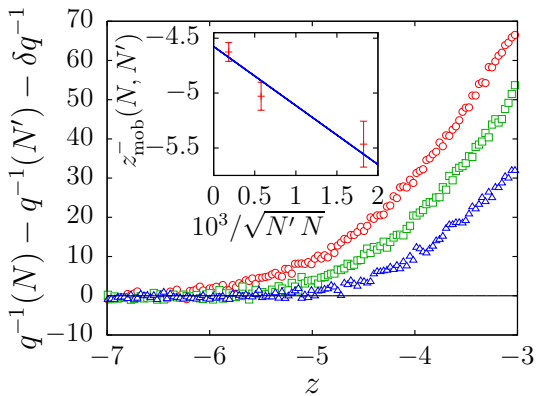}
\caption{An example of the procedure for establishing the mobility
  edge.
The symbols correspond to the  pairs of sizes $N=300$, $N'=1000$
(circles); $N=1000$, $N'=3000$ (squares); $N=3000$, $N'=10000$
(triangles). The estimated mobility edge for this pair is located
where the data fall below zero. In the inset, extrapolation of the
estimated mobility edge to infinite system size. 
}
\label{fig:randcubic-fitting-mobilityedge}
\end{figure}

\begin{figure}[t]
\includegraphics[scale=0.85]{%
\slaninafigdir/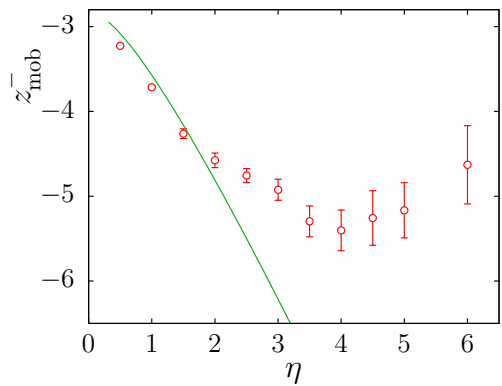}
\caption{Position of the mobility edge at the lower tail of the
  spectrum, for random cubic graph. The solid line is the band edge
  calculated in EMA.}
\label{fig:randcubic-mobilityedge-lower}
\end{figure}

We show in Fig. \ref{fig:randcubic-dos-1-2-3-4} the density of states
for several disorder strengths. The density of states is smooth and
free of singularities, which are typical of the spectrum of  ER graphs.
The localized states occur in the Lifschitz tails, as we can clearly
see in Figs. \ref{fig:randcubic-ipr-0.5-tail},
\ref{fig:randcubic-ipr-2-tail},
and \ref{fig:randcubic-ipr-4-tail}. Qualitatively, we observe that
localization is much stronger than in ER graphs and the IPR reaches
values very close to $1$. On the other hand, establishing the mobility edge 
is more difficult, because the deviations of the curves for different
$N$ are much smaller and obscured by statistical noise.
 We illustrate it in the inset of
Fig. \ref{fig:randcubic-ipr-4-tail}. In such a situation it is necessary
to develop a method for extracting the mobility edge as reliably as
possible. The method is illustrated in
Fig. \ref{fig:randcubic-fitting-mobilityedge}. The procedure we used
consists in comparing the difference 
of average IPR for two system sizes, $\Delta q^{-1}=q^{-1}(N)-q^{-1}(N')$ with the
level of statistical noise $\delta q^{-1}$. The estimate for the
mobility edge $z^-_\mathrm{mob}(N,N')$ is found where the difference $\Delta
q^{-1}$ as a function of $z$ crosses the noise level  $\delta
q^{-1}$. The error produced in this method is estimated in a similar
manner, as difference of points where $\Delta
q^{-1}(N,N')$ crosses  $\delta
q^{-1}$ and where it crosses twice as large noise $2\delta
q^{-1}$. The error bars shown if
Figs. \ref{fig:randmat-mobilityedge-lower} and
\ref{fig:randcubic-mobilityedge-lower} are obtained in this way.
We found that the estimate  $z^-_\mathrm{mob}(N,N')$ depends quite
strongly on the sizes $N$, $N'$. Therefore, we further
extrapolate the values found to infinite system, as shown in the inset
of Fig. \ref{fig:randcubic-fitting-mobilityedge}.

The dependence of the mobility edge on disorder strength is shown in
Fig. \ref{fig:randcubic-mobilityedge-lower}. As in the case of ER
graphs, we compare the dependence of the mobility edge on disorder
strength with the position of the band edge calculated using the
effective medium approximation. While in ER graph the EMA band edge
and the mobility edge go in parallel, in random cubic graph they
behave differently. While the EMA band edge grows in absolute value,
thus reflecting the overall broadening of the density of states for
increasing disorder, the mobility edge remains deep within the range of
the EMA band. For disorder stronger than about $\eta\simeq 4$ the
interval of extended states starts narrowing. This agrees
 qualitatively  with earlier results on Anderson
localization on Bethe lattice \cite{ab_tho_74,aiz_war_11} which state
that for strong enough disorder, $\eta>\eta_c$,
 all states are localized. Note that
the same qualitative behavior was also found by diagrammatic methods
for lattices in large Euclidean dimensions \cite{jan_kol_05}.

\begin{figure}[t]
\includegraphics[scale=0.85]{%
\slaninafigdir/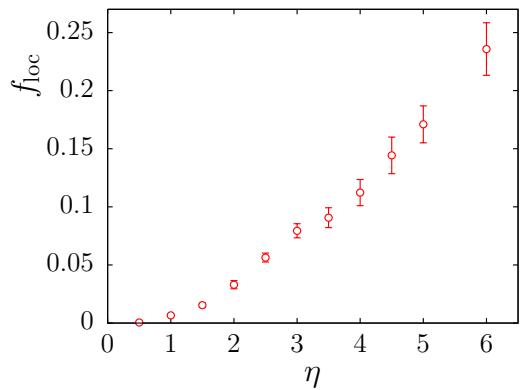}
\caption{Dependence of the fraction of states below the lower mobility edge
  on the strength of the disorder, for random cubic graph. The size is
  $N=10^4$.}
\label{fig:randcubic-fraction-localised}
\end{figure}

The fraction of states below the lower mobility edge is shown in
Fig. \ref{fig:randcubic-fraction-localised}. Again, the behavior is
completely different from the situation in ER graph. The fraction of
localized states is large and grows with the strength of the
disorder. We are unable to reach higher disorder strengths $\eta$,
because establishing the precise value of the mobility  edge is
increasingly difficult. However, our data are consistent with the
claim that beyond a critical strength of disorder the fraction reaches
its maximum, i. e. $f_\mathrm{loc}=1/2$ for $\eta>\eta_c$.

\subsection{IPR distribution}

\begin{figure}[t]
\includegraphics[scale=0.85]{%
\slaninafigdir/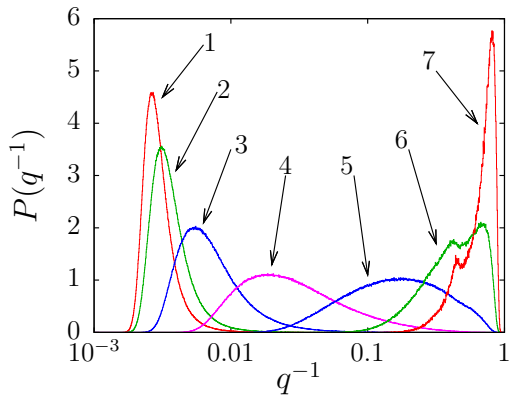}
\caption{Histogram of IPR, for states with eigenvalues within a fixed
  interval, for $\eta=2$ and $N=3000$. The arrows point to curves
  corresponding to intervals $z\in[-0.5,0.5]$ (line $1$), 
$[-1.5,-0.5]$ (line $2$), 
$[-2.5,-1.5]$ (line $3$), 
$[-3.5,-2.5]$ (line $4$), 
$[-4.5,-3.5]$ (line $5$), 
$[-5.5,-4.5]$ (line $6$), and
$[-6.5,-5.5]$ (line $7$). The data are accumulated from $17000$
  independent realizations.
}
\label{fig:randcubic-ipr-histogram-2-3000-allenergies}
\end{figure}

In addition to the dependence of the average IPR on $z$, we are interested
also in the fluctuations of IPR, if we restrict the eigenvalue to a
fixed interval $z\in[z_1,z_2]$. Indeed, we found that the fluctuations
may be very large, extending up to several orders of magnitude. We
show in Fig. \ref{fig:randcubic-ipr-histogram-2-3000-allenergies}
a series of histograms for the window $[z_1,z_2]$ sliding from
extended states through the transition region, to localized states. As
expected, the width of the distribution is largest around the
transition, but even in the localized regime it spans about one
decade.

\begin{figure}[t]
\includegraphics[scale=0.85]{%
\slaninafigdir/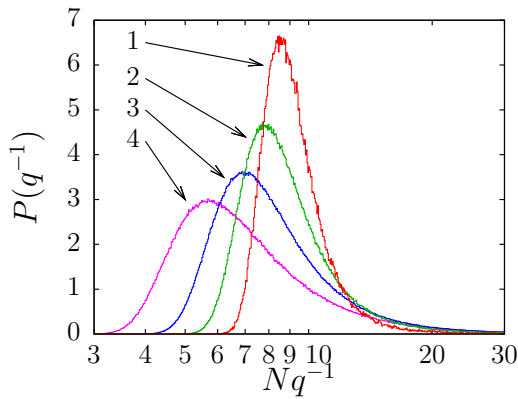}
\caption{Histogram of IPR in the range of extended states,
  $z\in[-0.1,0.1]$, for $\eta=2$ and different sizes of the system, $N=10^4$ (line
  $1$), $N=3000$ (line $2$), $N=1000$ (line $3$), and $N=300$ (line
  4). The data are accumulated from $610$, $17000$, $130000$, and
  $270000$ independent realizations, respectively.}
\label{fig:randcubic-ipr-histogram-2-deloc-allsizes}
\end{figure}

Let us first look at the extended states. The average IPR is expected
to scale as $1/N$. Therefore, we plot the histogram against the
rescaled value $Nq^{-1}$, in order to see the convergence for
increasing $N$. Indeed, we
can observe in Fig. \ref{fig:randcubic-ipr-histogram-2-deloc-allsizes}
that the position of the peak approaches to a limit and
simultaneously, the width of the peak shrinks. This suggests that
in the extended phase, IPR is a self-averaging quantity.

\begin{figure}[t]
\includegraphics[scale=0.85]{%
\slaninafigdir/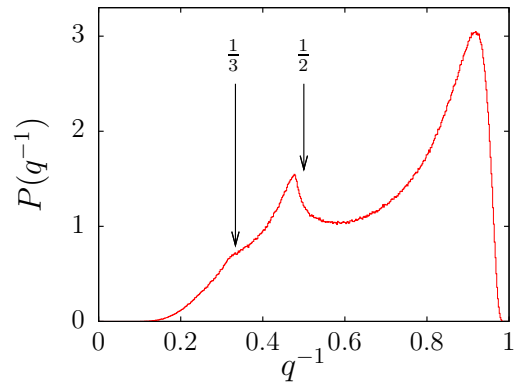}
\caption{Histogram of IPR in the range of localized states,
  $z\in[-8.5,-7.5]$, for $\eta=5$ and $N=1000$. The arrows indicate
  special values if IPR, $q^{-1}=1/2$ and $q^{-1}=1/3$. The data are
  accumulated from $280000$ independent realizations.}
\label{fig:randcubic-ipr-histogram-5-1000-loc}
\end{figure}

On the contrary, we found that in the localized phase the distribution
of IPR is independent of size. Moreover, as the example in
Fig. \ref{fig:randcubic-ipr-histogram-5-1000-loc} shows, there are
non-trivial structures in the distribution. In
Fig. \ref{fig:randcubic-ipr-histogram-5-1000-loc} we clearly see two
distinct peaks and a cusp. Interestingly, the positions of these three
structures are slightly below some special values of IPR, namely
$q^{-1}=1$, 
$q^{-1}=1/2$, and
$q^{-1}=1/3$. With our data available, 
we are unable to see further structures
at $q^{-1}=1/4$ etc., but we may speculate that they are also present.

\begin{figure}[t]
\includegraphics[scale=0.85]{%
\slaninafigdir/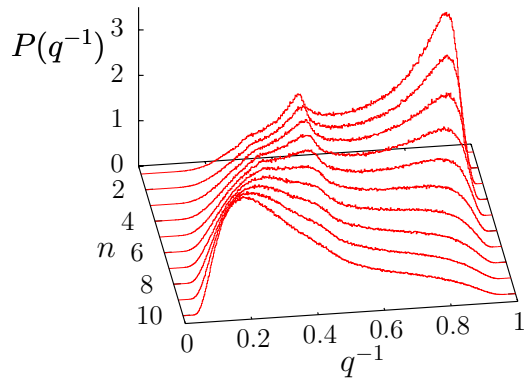}
\caption{Series of histograms of IPR in the range of localized states,
 for $\eta=5$ and $N=1000$. The index of the curve 
$n$ corresponds to the interval
 of eigenvalues according to the formula $z\in[-8.6+0.5n,-8.4+0.5n]$.
The data are
  accumulated from $280000$ independent realizations. }
\label{fig:randcubic-ipr-histogram-5-loc-allenergies}
\end{figure}

Further on, we want to see how these structures evolve when we sweep 
through the regime of
localized states, changing the value of $z$.
 We plot in
Fig. \ref{fig:randcubic-ipr-histogram-5-loc-allenergies} the series of
histograms for $z\in[-8.6+0.5n,-8.4+0.5n]$, $n=1,2,...,10$. For large
$|z|$, i. e. deep in the localized phase, the peak at $q^{-1}\simeq 1$
dominates, but when we decrease $|z|$, i. e. when we approach the
transition, the peak $q^{-1}\simeq 1/2$ takes over, and further on the
peak at $q^{-1}\simeq 1/3$ becomes most visible. Simultaneously the
peaks broaden and shift to lower values of IPR, so that the structure
of distinct peaks is less and less clear.

We can interpret the special positions of the peaks  at $q^{-1}=1$, 
$q^{-1}=1/2$, etc. as coming from eigenvectors localized mostly at
one, two, etc. sites. In order to support this interpretation, we
measured also the weighted average distance between sites. To this
end, we first find the shortest paths between each pair of vertices in
the current realization of the random cubic graph. Denote $d(i,j)$ the
length of this path for vertices $i$ and $j$. Of course, $d(i,i)=0$
for every $i$. Then, for each normalized eigenvector  $e_{i\lambda}$
we calculate the weighted average 
\begin{equation}
\overline{d}(\lambda)=\frac{\sum_{i,j=1}^N
  d(i,j)e^2_{i\lambda}e^2_{j\lambda}}{\sum_{i,j=1}^N
  e^2_{i\lambda}e^2_{j\lambda}}\;.
\end{equation}
For a vector strictly localized at one single site we get the average distance
$\overline{d}=0$, for a vector localized on a pair of neighbors it is
$\overline{d}=1/2$ and for a vector localized on a pair of sites at
distance $2$ we have $\overline{d}=1$. The two latter cases give the
same IPR, $q^{-1}=1/2$, so the average distance brings further
information on the eigenvector. We plot in
Fig. \ref{fig:randcubic-ipr-histogram-5-1000-loc} the joint
distribution of IPR and average distance, in the form of
two-dimensional histogram. The value of $P(q^{-1},\overline{d})$ is
discriminated by the color, higher values being darker. We clearly
observe two black spots corresponding to peaks of the
distribution. The first one is located about $q^{-1}\simeq 0.85$ and
$\overline{d}\simeq 0.2$, implying states localized around one single
site. The shift from the point $q^{-1}=1$, $\overline{d}=0$ is due to
decaying tails of the eigenvector. The second peak is slightly shifted
from the ideal position $q^{-1}=1/2$, $\overline{d}=1/2$. Clearly, it
corresponds to states localized on a pair of neighbors, again with
decaying tails. We can also see a darker spot around the position
$q^{-1}=1/2$, $\overline{d}=1$. This small peak indicates states
localized around a pair of sites at distance $2$, in. e. on second
neighbors.  

One might  rise a serious suspicion, that each of the peaks in the
histogram of IPR corresponds to different realization. If that were true,
the multi-peak structure would be the artifact of accumulating data
from many independent realizations into one histogram. To check it, we
calculated the same histogram for a large system, $N=30000$. In the
localized phase, we found
two distinct peaks also in the histogram for one single
realization. Moreover, comparing the histograms for a single realization and for
$20$ independent realizations, we see the same shape of the distribution,
within statistical errors. Therefore, the observed peculiarities in the
IPR distribution are characteristic of single realizations.

\begin{figure}[t]
\includegraphics[scale=0.85]{%
\slaninafigdir/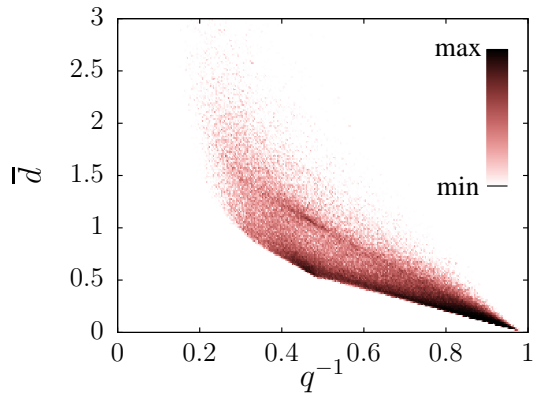}
\caption{Two-dimensional histogram of IPR and average distance of
  sites, in the range of localized states,
  $z\in[-7.1,-6.9]$, for $\eta=5$ and $N=1000$. Darker color indicates
  higher value of the histogram. The data are
  accumulated from $30000$ independent realizations.}
\label{fig:randcubic-ipr-dist-histogram-5-loc}
\end{figure}

\begin{figure}[t]
\includegraphics[scale=0.85]{%
\slaninafigdir/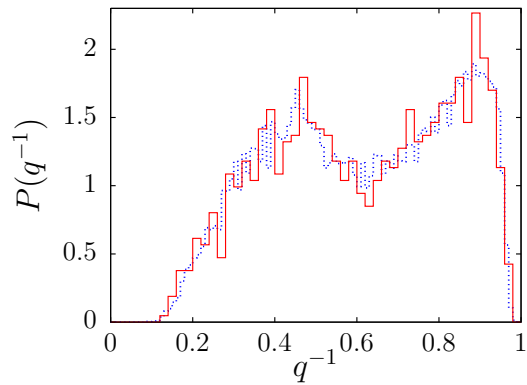}
\caption{Histogram of IPR in the range of localized states, $z\in[-7.5,6.5]$,
 for $\eta=6$ and $N=30000$. The solid line is the histogram for a
 single realization, while the dotted line is the cumulative histogram for
 $20$ independent realizations.}
\label{fig:randcubic-ipr-histogram-6-30000-loc}
\end{figure}

\subsection{Level spacings}

An important feature of the localization transition, stressed already
in the early works \cite{ab_an_tho_73,ab_tho_74}, is the qualitative change in 
 fluctuations of the imaginary part of the resolvent close to
 the real axis. It was used for establishing the mobility edge
 e. g. in Ref. \cite{cil_gri_mar_par_ver_05}. In fact, this feature is due to the change
 in level-spacing statistics \cite{eva_eco_92}. Extended states are supposed to obey the
 level-spacing distribution common to Gaussian orthogonal ensemble (GOE) of
 random matrices \cite{mehta_91}, i. e. in a very good approximation
\begin{equation}
P_\mathrm{GOE}(x)\propto x\mathrm{e}^{-x^2}\;.
\label{eq:goe-statistics}
\end{equation}
(In this expression $x$ is the distance of eigenvalues 
 normalized to the average level
spacing). On the other hand, localized states should obey the Poisson statistics
\begin{equation}
P_\mathrm{Poisson}(x)\propto\mathrm{e}^{-x}\;.
\label{eq:poisson-distribution}
\end{equation}
Intuitively, the change in statistics can be understood in terms of
level repulsion, which is substantial for extended, but very small for
localized states. Therefore, localized states behave as if they were
nearly independent and their energies scattered randomly, which gives
rise to the Poisson statistics. Because Poisson statistics is
characteristic for integrable systems, while statistics like
(\ref{eq:goe-statistics}) is the fingerprint of a chaotic system, the localization
transition can be viewed also as a chaotic-integrable transition.

\begin{figure}[t]
\includegraphics[scale=0.85]{%
\slaninafigdir/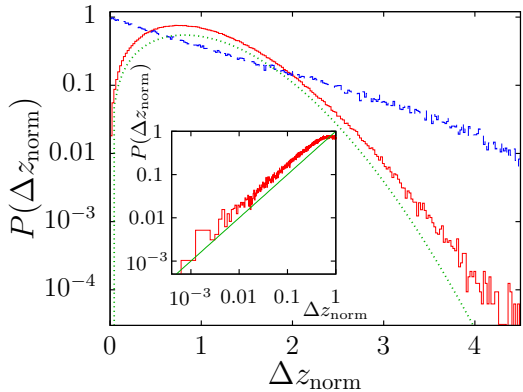}
\caption{Distribution of normalized level spacings in the spectrum of
  random cubic graph with disorder strength $\eta=2$ and size
  $N=1000$.
 The levels
  analyzed are restricted to intervals $z\in[-0.1,0.1]$ (solid line) 
and $z\in[-7,-6]$ (dashed line). The dotted line is the dependence
$\propto \Delta z_\mathrm{norm}\exp\big(-a(\Delta
z_\mathrm{norm})^2\big)$, with $a=0.75$, which corresponds to the Gaussian 
orthogonal ensemble. In the inset we show the detail of
the distribution at $z\in[-0.1,0.1]$ for very small spacings. The straight line is the
linear dependence $\propto\Delta z_\mathrm{norm}$.}
\label{fig:randcubic-levelspacing}
\end{figure}

We analyzed the random cubic graph of size $N=1000$ and disorder
strength $\eta=2$ and we extracted the level spacing statistics for
the spacings between
eigenvalues, normalized to the average spacing within certain
interval.
 We used the interval $z\in[-0.1,0.1]$ as a typical
representative of extended states and $z\in[-7,-6]$ as a representative
  of localized states. The results are shown in Fig.
  \ref{fig:randcubic-levelspacing}. The difference in statistics is
  clearly visible. The detail in the inset of Fig.
  \ref{fig:randcubic-levelspacing} shows also that the behavior for
  small level spacings is close to linear in the extended phase, in
  accord with Eq. (\ref{eq:goe-statistics}). We checked also that the
  distribution for localized states decays exponentially, as in 
Eq. (\ref{eq:poisson-distribution}). Thus, it is clearly demonstrated that
 the level spacing statistics  gets transformed from from Poisson to
  GOE  when we go from localized to extended regime in the spectrum.

\begin{figure}[t]
\includegraphics[scale=0.85]{%
\slaninafigdir/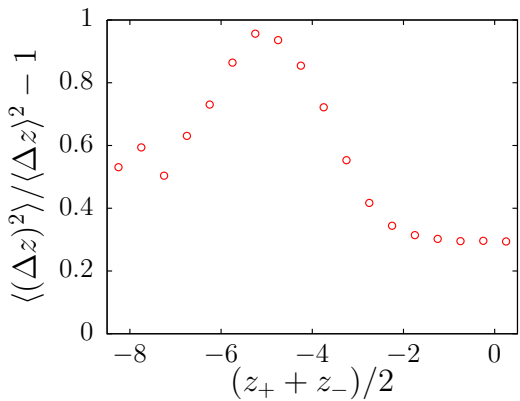}
\caption{Relative variance of the level spacing distribution,
  depending on the center of the interval over which the distribution
  is calculated. The  disorder strength is $\eta=2$ and the size
  of the system $N=1000$.
}
\label{fig:randcubic-levelspacing-moments}
\end{figure}

To make this argument quantitative, we calculate the moments of the
distribution of level spacings $\langle(\Delta z)^k\rangle=\int
(\Delta z)^k\,P(\Delta z)\mathrm{d}\Delta z$ within the interval
$z\in[z_-,z_+]$. Then, we plot in
Fig. \ref{fig:randcubic-levelspacing-moments} the relative variance of
the distribution $\langle (\Delta z)^2\rangle/\langle\Delta
z\rangle^2-1$. We can clearly see the peak around the transition
between localized and delocalized states, marking a qualitative change
in the level spacing distribution.

\section{Conclusions}

Numerically diagonalizing matrices up to size $10000\times 10000$, we
investigated localization transition in Erd\H os-R\'enyi and random
cubic graphs. In ER graphs, the free parameter was the average degree,
while in random cubic graphs, the parameter was the strength of disorder in the
diagonal matrix elements. The quantity to discriminate between localized and
extended regimes was the inverse participation ratio. We averaged IPR
over large number of realizations and using finite-size scaling, we
extracted the mobility edge. The benchmark for the position of the
mobility edge was the band edge found in the effective medium
approximation. 

The localization properties in ER and random cubic graphs are much
different. In the former, the mobility edge goes more or less in
parallel with the EMA band edge, 
when we change the average degree, and the fraction of
localized states decreases  when the average degree grows. In the latter, the EMA band
edge is significantly farther than the mobility edge, or else, much of
the localized states are actually present within the range of EMA
spectrum. The results are consistent with analytical findings which
predicted that a critical disorder strength exists, beyond which all
states are localized.

The inverse participation ratio exhibits rather strong
fluctuations. In the extended phase, the relative width of the IPR
distribution decreases with increasing system size, while in the 
localized phase the width of the distribution approaches a finite
value. Moreover, the distribution contains non-trivial structures of
several peaks. We interpret these structures as corresponding to
states localized around one, two, three, etc. sites.

For the random cubic graphs, we analyzed also the level spacing
statistics confirming the expectation that in the localized region
the statistics is close to Poissonian, while in the extended region it
is close to the statistics of Gaussian orthogonal ensemble.

\begin{acknowledgement}
I wish to thank to J. Ma\v{s}ek for numerous useful comments. 
This work was carried out within the project AV0Z10100520 of the Academy of 
Sciences of the Czech Republic and was 
supported by the M\v{S}MT of the Czech Republic, grant no. 
OC09078.

\end{acknowledgement}
\end{document}